\pdfoutput=1

\documentclass[11pt,dvipsnames]{article}

\usepackage[final]{acl}

\usepackage{times}
\usepackage{latexsym}

\usepackage[T1]{fontenc}

\usepackage[utf8]{inputenc}

\usepackage{microtype}

\usepackage{inconsolata}

\usepackage{pgfplots}
\pgfplotsset{compat=1.18}
\usepgfplotslibrary{colorbrewer}
\usetikzlibrary{backgrounds}
\usetikzlibrary{patterns}

\usepackage[inline]{enumitem}
\usepackage{booktabs}
\usepackage{amsmath}
\usepackage{amssymb}
\usepackage{multirow}

\newlength{\myl}
\usepackage{soul}

\newcommand{\emphh}[1]{\textcolor{WildStrawberry}{\textit{#1}}}

\definecolor{ori_color}{RGB}{219, 48, 122}
\definecolor{ours_color}{RGB}{119, 48, 122}
\definecolor{ours_norm_color}{RGB}{19, 48, 122}
\definecolor{sepbar}{RGB}{10, 40, 100}

\newcommand{\barc}[1]{\tikz[baseline=-.3em]{\draw (0,0) node[font=\scriptsize, text centered, rectangle, line width=.5pt, inner sep=2pt, draw=#1, fill=#1!50!white] {\phantom{1}};}}

\newcommand{\barr}{\tikz[baseline=-.3em]{\draw (0,0) node[font=\scriptsize,text centered, rectangle, line width=.5pt, inner sep=2pt, draw=ours_norm_color, fill=ours_norm_color!50!white, postaction={pattern=north east lines}] {\phantom{1}};}}

\title{An Analysis of Datasets, Metrics and Models in Keyphrase Generation}

\author{Florian Boudin \\
  JFLI, CNRS, Nantes University, France \\
  \texttt{florian.boudin@univ-nantes.fr} \\\And
  Akiko Aizawa \\
  National Institute of Informatics, Japan \\
  \texttt{aizawa@nii.ac.jp} \\}

\begin{document}
\maketitle
\begin{abstract}
Keyphrase generation refers to the task of producing a set of words or phrases that summarises the content of a document.
Continuous efforts have been dedicated to this task over the past few years, spreading across multiple lines of research, such as model architectures, data resources, and use-case scenarios.
Yet, the current state of keyphrase generation remains unknown as there has been no attempt to review and analyse previous work.
In this paper, we bridge this gap by presenting an analysis of over 50 research papers on keyphrase generation, offering a comprehensive overview of recent progress, limitations, and open challenges.
Our findings highlight several critical issues in current evaluation practices, such as the concerning similarity among commonly-used benchmark datasets and inconsistencies in metric calculations leading to overestimated performances.
Additionally, we address the limited availability of pre-trained models by releasing a strong PLM-based model for keyphrase generation as an effort to facilitate future research.
\end{abstract}

\section{Introduction}

Keyphrase generation involves generating a set of words or phrases that summarise the content of a source document.
These so-called keyphrases concisely and explicitly encapsulate the core content of a document, which makes them valuable for a variety of NLP and information retrieval tasks.
For instance, keyphrases were proven useful for improving document indexing~\cite{10.1145/42005.42016,zhai-1997-fast,10.1145/312624.312671,GUTWIN199981,boudin-etal-2020-keyphrase}, summarization~\cite{10.1145/564376.564398,wan-etal-2007-towards,10.1145/3442381.3449906,koto-etal-2022-lipkey} and question-answering~\cite{subramanian-etal-2018-neural,yang-etal-2019-sequential,lee-etal-2021-kpqa}, analyzing topic evolution~\cite{HU20191185,10.1007/s11192-020-03576-5,LU2021102594} or assisting with reading comprehension~\cite{10.5555/1769590.1769657,10.1002/asi.24749}.

Keyphrase generation expands on keyphrase extraction by enabling the production of \emph{keyphrases absent from the source text}~\cite{liu-etal-2011-automatic}.
This ability is critical when dealing with short documents that often lack appropriate keyphrase candidates.
%
%
%
%
\citet{meng-etal-2017-deep} provided the seminal work on keyphrase generation, introducing a sequence-to-sequence learning approach.
Their model builds upon an RNN encoder-decoder architecture~\cite{cho-etal-2014-learning,NIPS2014_a14ac55a} and incorporates a copying mechanism~\cite{gu-etal-2016-incorporating} to identify important phrases within the source text.
%
%
Equally importantly, they introduced KP20k, a dataset that laid the groundwork for end-to-end training of neural models for keyphrase generation.

%
%

Over the past few years, continuous efforts have been devoted to improve the effectiveness of keyphrase generation models.
These efforts have been spread across different lines of research, such as model architectures, data resources, and use-case scenarios, often pursued separately.
This analysis paper presents an overview of the current state of keyphrase generation, discussing recent progress, remaining limitations and open challenges.
More specifically, we compiled and analysed a collection of over 50 papers on keyphrase generation, identifying the type(s) of contribution these papers made (\S\ref{subsec:types-of-contribution}), examining the most frequently used benchmark datasets (\S\ref{subsec:benchmark-datasets}) and evaluation metrics (\S\ref{subsec:evaluation-metrics}), providing descriptions of proposed models while highlighting important milestones (\S\ref{subsec:proposed-models}), and investigating how proposed models perform against each other (\S\ref{subsec:empirical-results}).
%


Our findings are that: 1) commonly used benchmark datasets are so similar that reporting results on more than one adds no value, 2) the performance of models is often overestimated due to discrepancies in evaluation protocols, and 3) while dedicated models have been superseded by fine-tuned pre-trained language models (PLMs), the overall performance gain since early models remains limited.

Our work goes beyond surveying the existing literature and addresses some of the aforementioned concerns by training, documenting and releasing a strong PLM-based model for keyphrase generation along with an evaluation framework to facilitate future research (\S\ref{sec:sota-baseline-model}).
Finally, we discuss some of the open challenges in keyphrase generation and propose actionable directions to address them (\S\ref{sec:challenges}).

\section{Scope of the Study}

Our analysis encompasses a total of 52 research papers selected based on the following criteria: they are accessible through the ACL Anthology, they contain the phrase ``\textit{keyphrase generation}'' either in their titles or abstracts, and they have been published after the seminal work of~\citet{meng-etal-2017-deep}.
For a more comprehensive coverage, we also include papers from other NLP-related venues, comprising AAAI (4 papers), SIGIR (1 paper), and CIKM (1 paper).
To keep the number of papers manageable, we arbitrarily disregard papers from pre-print servers (e.g.~arXiv) or those published in non-ACL venues. 
%
%
Nonetheless, we are confident that our sample represents a comprehensive portion of the research on keyphrase generation, encompassing all papers published at major NLP venues in the last seven years. 
This includes, for instance, the ten most cited articles in the field.\footnote{\url{https://www.semanticscholar.org/search?q="keyphrase\%20generation"&sort=total-citations}}

For each paper in our sample, we manually collect the following information:
\begin{itemize}[topsep=0.3em,itemsep=0em]
    \item The \textbf{type(s) of contribution} the paper is making. 
    We adopt the ACL 2023 classification of contribution types~\cite{rogers-etal-2023-report}, which includes:
    \begin{enumerate*}[label=\arabic*)]
        \item NLP engineering experiment (most papers proposing methods to improve state-of-the-art),
        \item approaches for low-compute settings, efficiency,
        \item approaches for low-resource settings,
        \item data resources,
        \item data analysis,
        \item model analysis and interpretability,
        \item reproduction studies,
        \item position papers,
        \item surveys,
        \item theory,
        \item publicly available software and pre-trained models.
    \end{enumerate*}

    \item For papers proposing models, we record their \textbf{best scores} on each dataset they experiment with, in the form of $\langle dataset, metric, value \rangle$ triples.
    We extract scores primarily from the main tables of the content, supplementing with tables from appendices only if they report superior performance.
    In cases where multiple model variants are reported, we select the one demonstrating the best overall performance, or, when it is not clear, the one that performs best on the KP20k dataset.
    In total, we extracted 826 triples from our sample, corresponding to 50 distinct models.

    \item We also document the \textbf{architecture} of the proposed models (e.g~RNNs, Transformers), the use of \textbf{statistical significance tests} on the results, and the availability of both the \textbf{code} and the \textbf{model weights}. 
        
\end{itemize}

All the data collected in the course of this study is available at \url{https://github.com/boudinfl/kg-datasets-metrics-models}.

\section{Analysis}
\label{sec:analysis}

In this section, we analyze the selected papers across five key dimensions: types of contribution (\S\ref{subsec:types-of-contribution}), benchmark datasets (\S\ref{subsec:benchmark-datasets}), evaluation metrics (\S\ref{subsec:evaluation-metrics}), model architectures (\S\ref{subsec:proposed-models}), and the best reported performances (\S\ref{subsec:empirical-results}).

\subsection{Types of contribution}
\label{subsec:types-of-contribution}

We start our analysis by presenting statistics on the types of contribution made in the papers we examined (see Table~\ref{tab:contributions}).
Most of the papers propose new models (87\%), suggesting that the primary emphasis within the field is on improving the performance of the state-of-the-art.
This trend is reinforced by the fact that the second most common contribution is data resources (19\%), essential for validating improvements.
Some attention was given to model analysis and interpretability (14\%), particularly through empirical evaluations of multiple models~\cite{cano-bojar-2019-keyphrase,meng-etal-2021-empirical,meng-etal-2023-general,wu-etal-2023-rethinking-model} and evaluations via downstream tasks~\cite{boudin-etal-2020-keyphrase,boudin-gallina-2021-redefining}.
Approaches for low-resource settings also received some attention (14\%), initially with data-efficient models~\cite{lancioni-etal-2020-keyphrase,wu-etal-2022-representation}, then through data augmentation methods~\cite{gao-etal-2022-retrieval,garg-etal-2023-data,kang-shin-2024-improving} and most recently, domain adaptation strategies~\cite{boudin-aizawa-2024-unsupervised}.

One underexplored area is the development of low-compute approaches, maybe overlooked in the race toward larger models designed to boost performance.
This trend contrasts with practical applications, such as document indexing, where speed and efficiency are critical.
Our analysis also reveals the limited attention given to data analysis, reproduction studies and surveys in the literature.\footnote{We note, however, that several surveys on keyphrase extraction have been conducted; see Appendix~\ref{subsec:related-surveys} for a review.} 
This paper seeks to address this gap by providing new insights into the redundancy of existing datasets, conducting replication experiments on model evaluation, and offering a comprehensive overview of models for keyphrase generation.

\begin{table}[ht!]
    \centering
    \begin{tabular}{l|r}
\toprule
\textbf{Type of contribution} & \textbf{\%} \\
\midrule
NLP engineering experiment & 86.5 \\
Data resources & 19.2 \\
Model analysis and interpretability & 13.5 \\
Approaches for low-resource settings & 13.5 \\
Software and pre-trained models & 7.7 \\
Reproduction studies & 1.9 \\
\bottomrule
\end{tabular}

    \caption{Percentage of papers making each type of contribution (a paper may contribute to multiple types).}
    \label{tab:contributions}
\end{table}

\subsection{Benchmark Datasets}
\label{subsec:benchmark-datasets}

\begin{figure}[tb!]
    \centering
    \resizebox{\linewidth}{!}{%
    \pgfplotsset{compat=1.11,
    /pgfplots/ybar legend/.style={
    /pgfplots/legend image code/.code={%
       \draw[##1,/tikz/.cd,yshift=-0.25em]
        (0cm,0cm) rectangle (4pt,0.6em);},
   },
}

\begin{tikzpicture}
  \begin{axis}[
    ybar,
    legend style={legend columns=1,anchor=north east,draw=none,font=\footnotesize},
    legend cell align={left},
    width=10cm,
    height=5.5cm,
    ymin=0,
    ymax=46,
    xmin=-1,
    xmax=13,
    xtick style={draw=none},
    ytick pos=left, 
    ylabel={\# papers},
    xticklabels={\underline{KP20k},SemEval-2010,Inspec,Krapivin,NUS,\underline{KPTimes},DUC2001,\underline{StackEx (Yuan)},Weibo,StackEx (Wang),Twitter,JPTimes,\underline{OpenKP}},
    xtick={0,...,12},
    nodes near coords, 
	nodes near coords align={vertical},
    x tick label style={rotate=45, anchor=east, font=\small, yshift=-0.1cm, xshift=0.1cm} 
    ]

    \addplot[draw=RoyalPurple, fill=RoyalPurple!50!white, bar shift=0pt] coordinates { (0, 40) };
    \addlegendentry{Scientific abstracts}
    \addplot[draw=RoyalPurple, fill=RoyalPurple!50!white, bar shift=0pt, forget plot] coordinates { (1, 35) };
    \addplot[draw=RoyalPurple, fill=RoyalPurple!50!white, bar shift=0pt, forget plot] coordinates { (2, 35) };
    \addplot[draw=RoyalPurple, fill=RoyalPurple!50!white, bar shift=0pt, forget plot] coordinates { (3, 32) };
    \addplot[draw=RoyalPurple, fill=RoyalPurple!50!white, bar shift=0pt, forget plot] coordinates { (4, 32) };
    \addplot[draw=PineGreen, fill=PineGreen!50!white, bar shift=0pt] coordinates { (5, 9) };
    \addlegendentry{News}
    \addplot[draw=PineGreen, fill=PineGreen!50!white, bar shift=0pt, forget plot] coordinates { (6, 8) };
    \addplot[draw=WildStrawberry, fill=WildStrawberry!50!white, bar shift=0pt] coordinates { (7, 4) };
    \addlegendentry{Social media}
    \addplot[draw=WildStrawberry, fill=WildStrawberry!50!white, bar shift=0pt, forget plot] coordinates { (8, 3) };
    \addplot[draw=WildStrawberry, fill=WildStrawberry!50!white, bar shift=0pt, forget plot] coordinates { (9, 3) };\addplot[draw=WildStrawberry, fill=WildStrawberry!50!white, bar shift=0pt, forget plot] coordinates { (10, 2) };\addplot[draw=PineGreen, fill=PineGreen!50!white, bar shift=0pt, forget plot] coordinates { (11, 2) };
    \addplot[draw=GreenYellow, fill=GreenYellow!50!white, bar shift=0pt] coordinates { (12, 2) };
    \addlegendentry{Web pages}

    
  \end{axis}
\end{tikzpicture}


    }
    \caption{Number of papers utilizing each dataset. \underline{Underlined datasets} contain $100k$+ training samples. Datasets used only once are omitted for clarity. }
    \label{fig:datasets}
\end{figure}
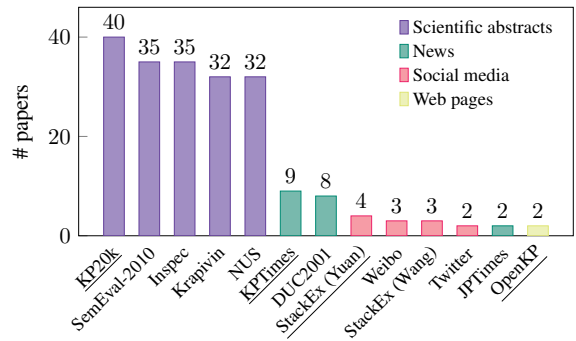

We proceed with our analysis by examining the most frequently used datasets (see Figure~\ref{fig:datasets}, detailed statistics of the datasets are provided in \S\ref{sec:details-datasets}).
We find that 26 distinct datasets were employed across the examined papers, with five datasets notably more prevalent than others: KP20k~\cite{meng-etal-2017-deep}, SemEval-2010~\cite{kim-etal-2010-semeval}, Inspec~\cite{hulth-2003-improved}, Krapivin~\cite{krapivin2009large}, and NUS~\cite{10.1007/978-3-540-77094-7_41}.
These datasets are commonly used together, with 22 out of 52 papers (42\%) employing all five, and 39 out of 52 (75\%) employing at least two.
All five datasets exclusively contain scientific abstracts, whereas the remaining datasets are sourced from various domains, such as news, social media and web pages.
This domain bias can be attributed to two main factors: the availability of scientific abstracts, and the frequent presence of author-assigned keyphrases, serving as naturally occurring ground truth.
When considering size, only a handful of datasets contain a sufficient number of samples (i.e.~$>100k$ training samples, underlined in Figure~\ref{fig:datasets}) to effectively train generative models.
The majority of these datasets, however, are relatively small (i.e.~$<1k$ samples) and are mainly used for testing purposes.

A closer examination of the five widely-used datasets reveals substantial overlap.
All consist of scientific abstracts from the Computer Science domain, and at least three---KP20k, SemEval-2010, and Krapivin---share documents from the same source, the ACM Digital Library.
This raises concerns about potential data leakage and questions the value of using these datasets together in experimental setups.
%
%
%

\begin{figure}[hb!]
    \centering
    \includegraphics[width=\linewidth]{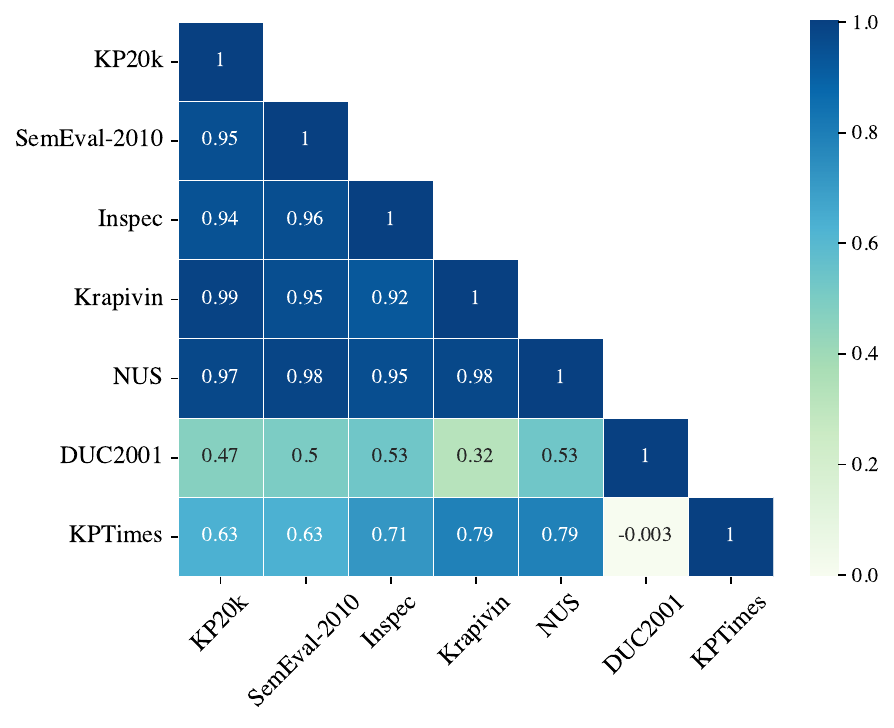}
    \caption{Pearson's correlation coefficient $\rho$ computed between the model scores across datasets.}
    \label{fig:correlations}
\end{figure}

To shed light on these questions, we measured the correlation between the model scores across datasets, exploring whether models perform uniformly across different datasets.
Our objective here is to determine the extent to which including more than one of these datasets in the experiments of a paper provides additional insights.
From the correlation matrix in Figure~\ref{fig:correlations}, we see that the performance of models among the five widely-used datasets is almost perfectly correlated (Pearson's correlation coefficient $\rho> 0.9$, $\text{p-value} < 0.01$).
This observation implies that \emphh{there is no practical benefit in reporting the results on more than one of these five datasets}, despite the common practice among previous studies of doing so.


\subsection{Evaluation Metrics}
\label{subsec:evaluation-metrics}

\begin{figure}[htb!]
    \centering
    \resizebox{\linewidth}{!}{%
    \pgfplotsset{compat=1.11,
    /pgfplots/ybar legend/.style={
    /pgfplots/legend image code/.code={%
       \draw[##1,/tikz/.cd,yshift=-0.25em]
        (0cm,0cm) rectangle (4pt,0.6em);},
   },
}

\begin{tikzpicture}
  \begin{axis}[
    ybar,
    legend style={legend columns=1,anchor=north east,draw=none,font=\footnotesize},
    legend cell align={left},
    width=10cm,
    height=5.5cm,
    ymin=0,
    ymax=40,
    xmin=-1,
    xmax=13,
    xtick style={draw=none},
    ytick pos=left, 
    ylabel={\# papers},
    xticklabels={},
    xtick={0,...,13},
    nodes near coords, 
	nodes near coords align={vertical},
    xticklabels={$F_1@5$,$F_1@M$,$F_1@M$,$F_1@5$,$F_1@10$,$R@10$,$R@50$,$F_1@O$,$F_1@10$,$F_1@5$,$F_1@M$, $F_1@3$, $mAP$},
    x tick label style={rotate=45, anchor=east, font=\small, yshift=-0.1cm, xshift=0.1cm} 
    ]

    \addplot[draw=RoyalPurple, fill=RoyalPurple!50!white, bar shift=0pt] coordinates { (0, 34) };
    \addlegendentry{Present}
    \addplot[draw=RoyalPurple, fill=RoyalPurple!50!white, bar shift=0pt, forget plot] coordinates { (1, 28) };
    \addplot[draw=PineGreen, fill=PineGreen!50!white, bar shift=0pt] coordinates { (2, 25) };
    \addlegendentry{Absent}
    \addplot[draw=PineGreen, fill=PineGreen!50!white, bar shift=0pt, forget plot] coordinates { (3, 22) };
    \addplot[draw=RoyalPurple, fill=RoyalPurple!50!white, bar shift=0pt, forget plot] coordinates { (4, 13) };
    \addplot[draw=PineGreen, fill=PineGreen!50!white, bar shift=0pt, forget plot] coordinates { (5, 11) };
    \addplot[draw=PineGreen, fill=PineGreen!50!white, bar shift=0pt, forget plot] coordinates { (6, 7) };
    \addplot[draw=RoyalPurple, fill=RoyalPurple!50!white, bar shift=0pt, forget plot] coordinates { (7, 5) };
    \addplot[draw=WildStrawberry, fill=WildStrawberry!50!white, bar shift=0pt] coordinates { (8, 5) };
    \addlegendentry{Combined}
    \addplot[draw=WildStrawberry, fill=WildStrawberry!50!white, bar shift=0pt, forget plot] coordinates { (9, 5) };
    \addplot[draw=WildStrawberry, fill=WildStrawberry!50!white, bar shift=0pt, forget plot] coordinates { (10, 4) };
    \addplot[draw=WildStrawberry, fill=WildStrawberry!50!white, bar shift=0pt, forget plot] coordinates { (11, 3) };
    \addplot[draw=WildStrawberry, fill=WildStrawberry!50!white, bar shift=0pt, forget plot] coordinates { (12, 3) };


    
  \end{axis}
\end{tikzpicture}
    }
    \caption{Number of papers employing each evaluation metric. Metrics used < 3 times are excluded for clarity. }
    \label{fig:metrics}
\end{figure}
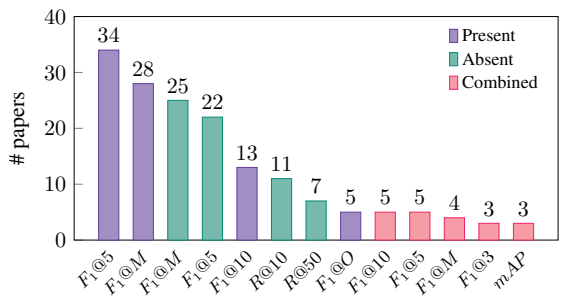

We move forward with our analysis by examining the evaluation of automatically generated keyphrases within our sample of papers.
With the exception of~\cite{Wu_Ma_Liu_Chen_Nie_2022}, all the proposed models are solely assessed through intrinsic evaluation, which involves comparing their output with a single ground truth, typically using exact matching.
%
%
From the extracted score triples, we find that 42 distinct evaluation metrics were reported across the papers (see Figure~\ref{fig:metrics}, detailed definitions of the evaluation metrics are provided in \S\ref{sec:details-eval}).
The majority of papers describing models (40 out of 50, 80\%) provide separate results for present and absent keyphrases, following the methodology of~\cite{meng-etal-2017-deep}.
As for the metrics, there is a high degree of consensus on the $F_1$ measure, with two configurations standing out: $F_1@M$ (using all the keyphrases predicted by the model) and $F_1@k$ (using the top-$k$ predicted keyphrases, with $k \in \{5, 10\}$).
%
%
%

Upon closer inspection of the evaluation settings in our sample of papers, we identified two major inconsistencies in how metrics are calculated.
First, two variants of $F_1@k$ co-exist.
%
%
%
Starting with \cite{chan-etal-2019-neural}, model predictions that do not reach $k$ keyphrases are extended with incorrect (dummy) phrases.
This prevents $F_1@k$ and $F_1@M$ scores from being identical, but lowers the scores for models generating fewer than $k$ keyphrases.
More critically, this practice undermines direct comparability with earlier work.

Second, we find that some form of normalization procedure is frequently applied prior to computing evaluation metrics, as observed in at least 30 out of 50 papers (60\%).\footnote{This information is often difficult to locate, as it is frequently omitted from papers and requires examining the source code and data.}
This procedure, commonly referred to as~\citet{meng-etal-2017-deep}'s pre-processing, is applied to ground-truth keyphrases and involves the following steps: 1) removing all the abbreviations/acronyms in parentheses, 2) tokenizing on non-letter characters, and 3) replacing digits with symbol \texttt{<digit>}.
This normalization impacts the evaluation (see an example in Table~\ref{tab:example-normalization} in \S\ref{subsec:example-normalization}), potentially leading to an overestimation of model performance and jeopardizing comparability with studies that do not employ it.

To gain insights on this issue, we conducted a series of replication experiments by reassessing the performance of three models---catSeqTG-2RF1~\cite{chan-etal-2019-neural}, ExHiRD-h~\cite{chen-etal-2020-exclusive} and SetTrans~\cite{ye-etal-2021-one2set}---for which the authors stated that they applied this normalization and provided the outputs of their model.
%
To ensure comparability and consistency, we compute $F_1@k$ scores with dummy phrases when the number of predicted keyphrases is less than $k$, following \cite{chan-etal-2019-neural} and subsequent works.

\begin{figure}[b!]
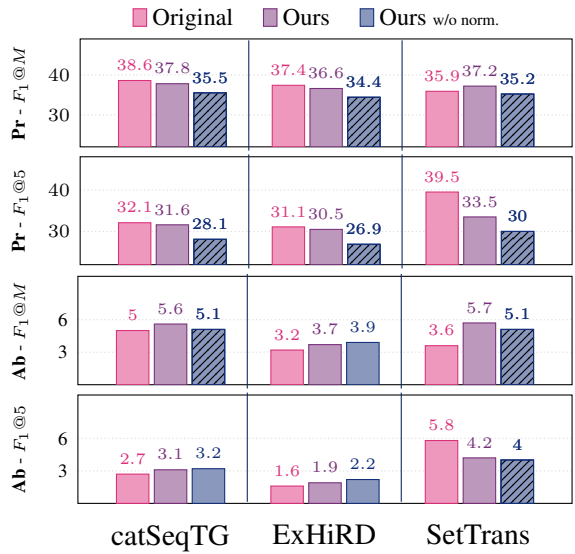

    \raggedleft
    \resizebox{\linewidth}{!}{
\begin{tikzpicture}[tight background]
    \input{figures/measure/pres-fm}
\end{tikzpicture}
}

\resizebox{\linewidth}{!}{
\begin{tikzpicture}[tight background]
    \input{figures/measure/pres-f5}
\end{tikzpicture}
}

\resizebox{\linewidth}{!}{
\begin{tikzpicture}[tight background]
    \input{figures/measure/abs-fm}
\end{tikzpicture}
}

\resizebox{\linewidth}{!}{
\begin{tikzpicture}[tight background]
    \input{figures/measure/abs-f5}
\end{tikzpicture}
}

    \caption{Replicated evaluation results on the KP20k dataset, alongside the performance reported in the original paper. Dashed bars (\,\protect\barr{}\,) indicate a significant decrease of performance compared to normalization, as determined by the Student's paired t-test ($\text{p-value} < 0.01$).}
    \label{fig:eval_replication}
\end{figure}

From the results in Figure~\ref{fig:eval_replication}, we observe that applying the normalization procedure significantly increases the scores for the majority of the evaluation metrics.
The impact of the normalization procedure is more pronounced for present keyphrases, showing an absolute difference of +2.2 points ($F_1@M$) and +3.5 points ($F_1@5$).
We notice a some difference in scores between the original (\,\barc{ori_color}\,) and our replicated evaluation (\,\barc{ours_color}\,), which we attribute to our method for determining whether a keyphrase is present or absent in the source document (see \ref{sec:details-eval}).
These observations alert that \emphh{the performance of many models have been overestimated from using this normalization procedure}, advocating for a cautious comparison of results between studies.
%






\subsection{Proposed Models}
\label{subsec:proposed-models}

In this section, we take a closer look at the models proposed in our sample of papers.
Figure~\ref{fig:tree} presents an overview of these models in the form of an evolutionary tree, highlighting five works that we consider important milestones for keyphrase generation.
In short, we first witness early efforts dedicated to refining the task formulation of keyphrase generation, followed by a transitional phase from RNN-based to Transformers-based models, and most recently, the adoption of fine-tuned PLMs.
Below, we provide brief descriptions of each model, organized around these milestone works and presented in chronological order.
\begin{itemize}[topsep=0.3em,itemsep=0.3em,leftmargin=2.3em,font=\bfseries\small]
    \item[2017] \citet{meng-etal-2017-deep} introduced a \emph{RNN-based encoder-decoder model for keyphrase generation}, alongside the KP20k dataset. This model was further improved with additional decoding mechanisms~\cite{chen-etal-2018-keyphrase,zhao-zhang-2019-incorporating}, multi-task learning~\cite{ye-wang-2018-semi}, external resources~\cite{chen-etal-2019-integrated}, latent topic information~\cite{wang-etal-2019-topic-aware,10.1145/3477495.3531990}, better encoding techniques~\cite{Chen_Gao_Zhang_King_Lyu_2019,kim-etal-2021-structure}, or self-training~\cite{Shen_Wang_Meng_Shang_2022}.

\begin{figure}[t!]
    \centering
    \resizebox{\linewidth}{!}{%
    \includegraphics{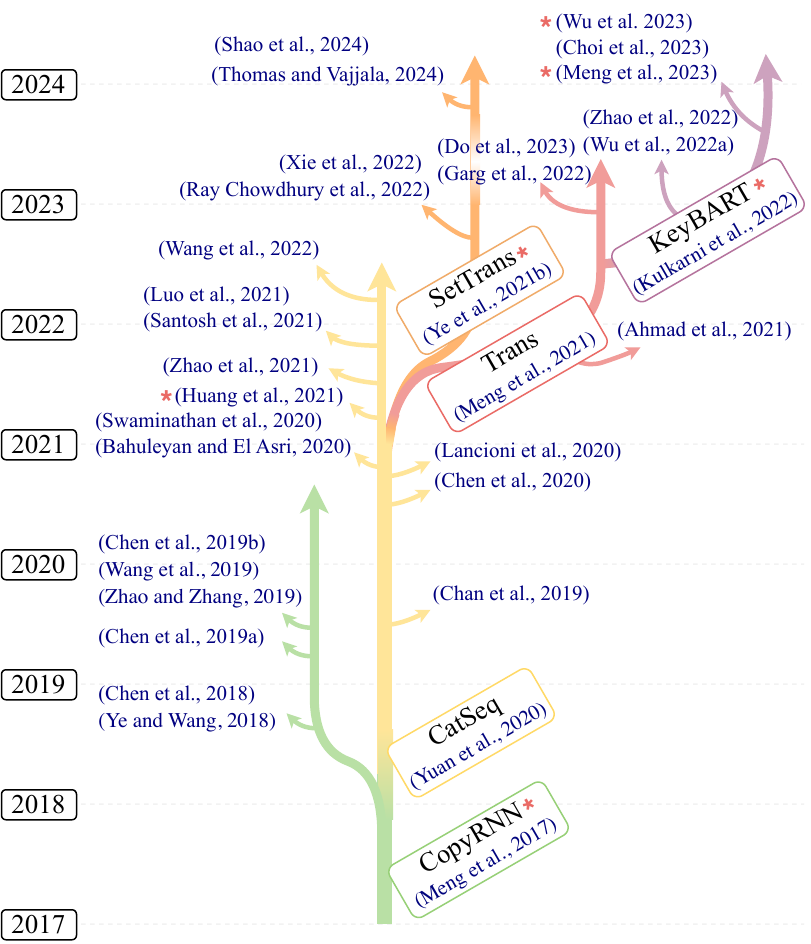}
    }
    \caption{Evolutionary tree of the keyphrase generation models in our analysis. Some models are omitted for clarity. \textcolor{red}{$^*$} indicate that the model weights are available.}
    \label{fig:tree}
\end{figure}

    \item[2018] \citet{yuan-etal-2020-one} introduced the \textsc{one2many} \emph{training paradigm}, enabling models to generate a variable number of keyphrases.\footnote{This work was submitted to arXiv in October 2018.} Subsequent studies have improved upon this work through the use of reinforcement learning~\cite{chan-etal-2019-neural,luo-etal-2021-keyphrase-generation}, hierarchical decoding~\cite{chen-etal-2020-exclusive}, GANs~\cite{lancioni-etal-2020-keyphrase,swaminathan-etal-2020-preliminary}, diversity promotion~\cite{bahuleyan-el-asri-2020-diverse}, or diverse decoding strategies~\cite{Huang_Xu_Jiao_Zu_Zhang_2021,zhao-etal-2021-sgg,10.1145/3459637.3482119,wang-etal-2022-automatic-keyphrase}.

    \item[2021] \citet{meng-etal-2021-empirical} explored the generalization capabilities of keyphrase generation models and were among the first to apply \emph{Transformers for this task}.
    Other works improved the performance of Transformers-based models though manipulation of the input document~\cite{ahmad-etal-2021-select,garg-etal-2022-keyphrase} or guided decoding~\cite{do-etal-2023-unsupervised}.

    \item[2021] \citet{ye-etal-2021-one2set} proposed the \textsc{one2set} \emph{training paradigm} that utilizes control codes to generate a set of keyphrases. Further work improved this approach through data augmentation~\cite{ray-chowdhury-etal-2022-kpdrop}, model calibration~\cite{xie-etal-2022-wr}, joint keyphrase extraction~\cite{thomas-vajjala-2024-improving} or LLM verification~\cite{shao-etal-2024-one2set}.

    \item[2022] \citet{kulkarni-etal-2022-learning} investigated the utilization of \emph{PLMs for keyphrase generation}.
    Subsequent studies confirmed that fine-tuning a PLM, namely BART~\cite{lewis-etal-2020-bart}, for keyphrase generation achieves SOTA results~\cite{wu-etal-2021-unikeyphrase,houbre-etal-2022-large,wu-etal-2022-representation,meng-etal-2023-general,wu-etal-2023-rethinking-model}, and further improved its performance through output filtering~\cite{zhao-etal-2022-keyphrase}, low-resource fine-tuning~\cite{wu-etal-2022-representation,kang-shin-2024-improving,boudin-aizawa-2024-unsupervised}, contrastive learning~\cite{choi-etal-2023-simckp} or encoder-only models~\cite{wu-etal-2024-leveraging}.
    
\end{itemize}

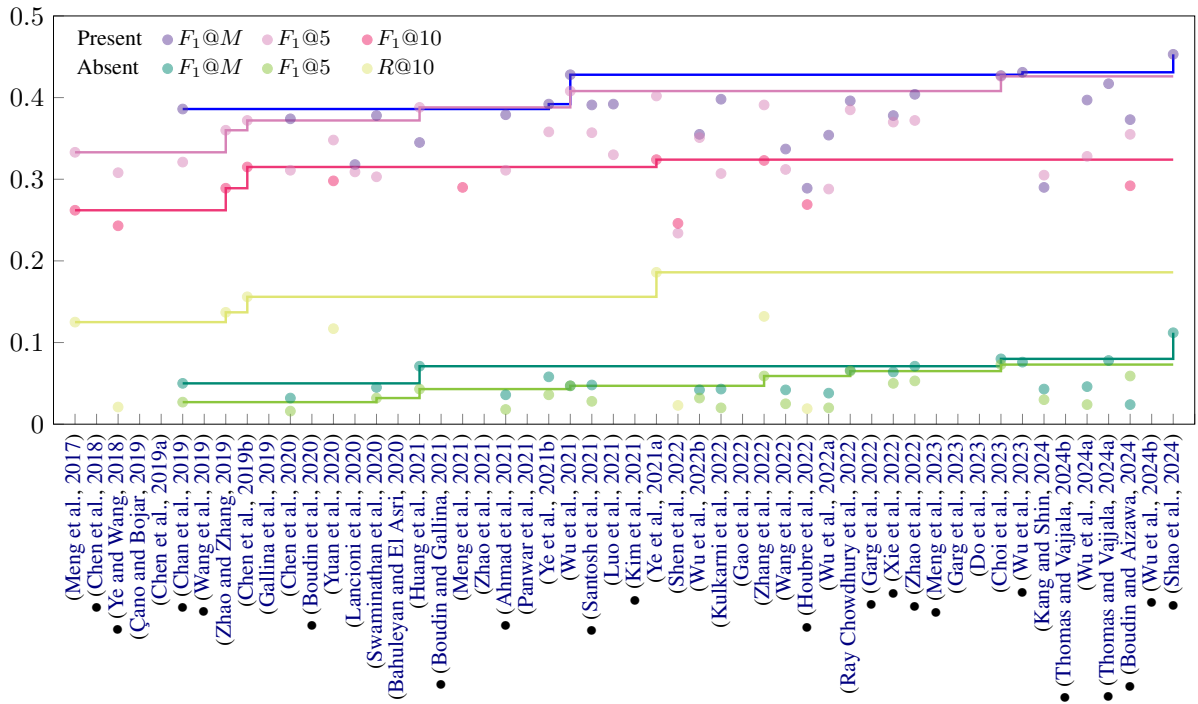
\begin{figure*}[ht!]
    \centering
    \resizebox{\textwidth}{!}{%
    \pgfplotstableread{figures/data/kp20k.dat}\datatable

\begin{tikzpicture}
\begin{axis}[
    ymin=0,
    ymax=0.5,
    xmin=-1,
    xmax=52,
    width=18cm,
    height=7.5cm,
    xtick pos=bottom,
    ytick pos=left, 
    xtick={0,...,52},
    xticklabels from table={\datatable}{bibkey},
    x tick label style={rotate=90,anchor=east,font=\small},
    legend columns=4,
    legend style={at={(0.01,0.99)},anchor=north west},
    legend style={font=\footnotesize, draw=none, text width=3em},
    ]

    \addlegendimage{empty legend}
    \addlegendentry{Present}
    
    \addplot[opacity=0.5, only marks, color=RoyalPurple, mark=*] table [x=number,y=F1@M-present] {figures/data/kp20k.dat};
    \addlegendentry{$F_1@M$~~}
    \addplot[const plot, forget plot, mark=none, color=blue, line width=1pt ] table [x=number,y=bestF1@M-present] {\datatable};

    \addplot[opacity=0.5, only marks, color=Thistle, mark=*] table [x=number,y=F1@5-present] {figures/data/kp20k.dat};
    \addlegendentry{$F_1@5$}
    \addplot[forget plot, const plot, mark=none, color=Thistle, line width=1pt] table [x=number,y=bestF1@5-present] {\datatable};

    \addplot[opacity=0.5,only marks, color=WildStrawberry, mark=*] table [x=number,y=F1@10-present] {figures/data/kp20k.dat};
    \addlegendentry{$F_1@10$}
    \addplot[opacity=0.9, forget plot, const plot, mark=none, color=WildStrawberry, line width=1pt] table [x=number,y=bestF1@10-present] {\datatable};

    \addlegendimage{empty legend}
    \addlegendentry{Absent}

    \addplot[opacity=0.5, only marks, color=PineGreen, mark=*] table [x=number,y=F1@M-absent] {figures/data/kp20k.dat};
    \addlegendentry{$F_1@M$}
    \addplot+[const plot, forget plot, mark=none, color=PineGreen, line width=1pt] table [x=number,y=bestF1@M-absent] {\datatable};

    \addplot[opacity=0.5, only marks, color=LimeGreen, mark=*] table [x=number,y=F1@5-absent] {figures/data/kp20k.dat};
    \addlegendentry{$F_1@5$}
    \addplot[forget plot, const plot, mark=none, color=LimeGreen, line width=1pt] table [x=number,y=bestF1@5-absent] {\datatable};

    \addplot[opacity=0.5, only marks, color=GreenYellow, mark=*] table [x=number,y=R@10-absent] {figures/data/kp20k.dat};
    \addlegendentry{$R@10$}
    \addplot[forget plot, const plot, mark=none, color=GreenYellow, line width=1pt] table [x=number,y=bestR@10-absent] {\datatable};



\end{axis}

\end{tikzpicture}
    }
    \caption{Best scores achieved by each model in terms of $F_1@M$, $F_1@5$ and $F_1@10$ for present keyphrases and $F_1@M$, $F_1@5$ and $R@10$ for absent keyphrases on the KP20k dataset. The lines represent the state-of-the-art performance over time. $\bullet$ indicate that the paper utilizes statistical tests to validate the significance of the results.}
    \label{fig:sota-performance}
\end{figure*}

\begin{figure}[hb!]
    \centering
    \pgfplotstableread{figures/data/architectures.dat}\architable
    
\begin{tikzpicture}
\begin{axis}[
        ybar stacked,
        ymin=0,
        ymax=15,
        width=\linewidth,
        height=4cm,
        xtick={17,...,24},
        xtick style={draw=none},
        ytick pos=left, 
        ylabel={\# models},
        y tick label style={font=\small},
        every node near coord/.append style={font=\footnotesize},
        nodes near coords,
        coordinate style/.condition={meta < 2}{
            opacity=0,
        },
        legend style={at={(0.01,0.99)},legend columns=1,anchor=north west,draw=none,font=\footnotesize},
        legend cell align={left},
    ]

    \addplot [draw=PineGreen, fill=PineGreen!50!white] table [x=year,y=rnn] {\architable};
    \addlegendentry{RNN}
    
    \addplot [draw=Thistle, fill=Thistle!50!white] table [x=year,y=transformers] {\architable};
    \addlegendentry{Transformers}
\end{axis}
\end{tikzpicture}
    \caption{Architectures of the proposed keyphrase generation models over the years.}
    \label{fig:architectures}
\end{figure}
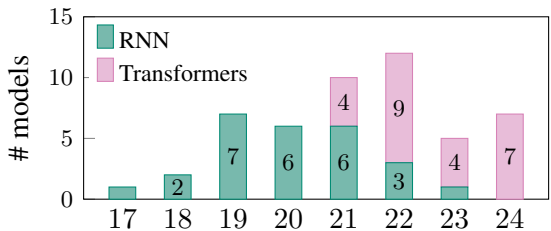

Figure~\ref{fig:architectures} provides a more detailed depiction of the architectures used by the proposed keyphrase generation models over the years.
Starting from 2021, we observe a swift transition from RNNs to Transformers, accelerated by the recent line of research on fine-tuning PLMs for the task.
This trend aligns with observations across numerous other NLP tasks, where (pre-trained) Transformers consistently achieve state-of-the-art performance.

While it is relatively common for studies introducing models to release the code for reproducing their experiments (34 out of 50, 68\%), it is much rarer for the model weights to be made available, with only 8 out of 50 studies doing so (marked with the symbol \textcolor{red}{$*$} in Figure~\ref{fig:tree}).
Importantly, code availability alone is not enough for reproducing the results reported in published literature~\cite{arvan-etal-2022-reproducibility-computational}.
%
This lack of model weights complicates fair comparisons between models and imposes unnecessary computational and environmental costs associated with retraining.
%

\subsection{Empirical Results}
\label{subsec:empirical-results}

We conclude our analysis by conducting a large-scale comparison of the performance of the proposed models in our sample of papers, focusing on the best scores they achieve on the KP20k benchmark dataset (see Figure~\ref{fig:sota-performance}). 
We plot the state-of-the-art performance over time using lines, considering the three most commonly used evaluation metrics for both present and absent keyphrases.
To the best of our knowledge, this is the comprehensive overview of state-of-the-art keyphrase generation model performance over time.

Overall, we observe a modest yet steady increase in state-of-the-art performance, with the most recent leap attributed to the use of LLMs for filtering keyphrase candidates generated by a fine-tuned PLM~\cite{shao-etal-2024-one2set}.
Two additional observations can be gleaned from the Figure: 1)~the absolute improvement in state-of-the-art performance since earlier works is limited; for instance, only 3.1\% in present $F_1@M$ separates the works of~\citet{chan-etal-2019-neural} and \citet{thomas-vajjala-2024-improving}; and 2)~the performance in absent keyphrase prediction remains notably low, barely reaching 11\% in $F_1@M$.
For the latter, we believe that the reasons could be traced back to the unreliability of the evaluation metrics, which  rely on strict matching against a single ground truth (see \S\ref{subsec:evaluation-metrics}).
This issue becomes more pronounced in the case of absent keyphrases where lexical variation is more prevalent, leading to lower scores.

Another notable observation is the limited use of statistical significance testing in the results, with only 20 out of the 50 doing so (marked with $\bullet$ in Figure~\ref{fig:sota-performance}).
We assume this is a consequence of the scarce availability of model weights (see \S\ref{subsec:proposed-models}), which hinders the reproducibility of prior research and the ability to directly compare model outputs.
Yet, statistical significance testing is crucial to assess the likelihood of potential improvements to models occurring by chance~\cite{dror-etal-2018-hitchhikers}, casting doubts on the actual progress of the task.

\section{A strong baseline model}
\label{sec:sota-baseline-model}

Our analysis highlights the progress achieved by current keyphrase generation models, while drawing attention to the lack of standardized evaluation procedures and the limited availability of pre-trained models.
To address these challenges, we provide a strong baseline model for keyphrase generation, along with an evaluation framework to facilitate future research.

Upon analysing the proposed model scores (see \S\ref{subsec:empirical-results}), we find that fine-tuning a PLM for the task consistently yields the best performance.
Based on this observation, we adopt this approach for our baseline model, leveraging BART-large~\cite{lewis-etal-2020-bart} as the initial PLM, in line with recent studies~\cite{meng-etal-2023-general,wu-etal-2023-rethinking-model}.
The model is fine-tuned on the KP20k training set for 10 epochs in a \textsc{One2Many} setting~\citep{yuan-etal-2020-one}, that is, given a source text as input, the task is to generate keyphrases as a single sequence of delimiter-separated phrases.
During fine-tuning, gold keyphrases are arranged in the present-absent order which was found to give the best results~\cite{meng-etal-2021-empirical}.
%
%
Notably, we do not apply any pre-processing to either the source texts or the ground-truth keyphrases, thereby fixing the issues we identified in \S\ref{subsec:evaluation-metrics}.

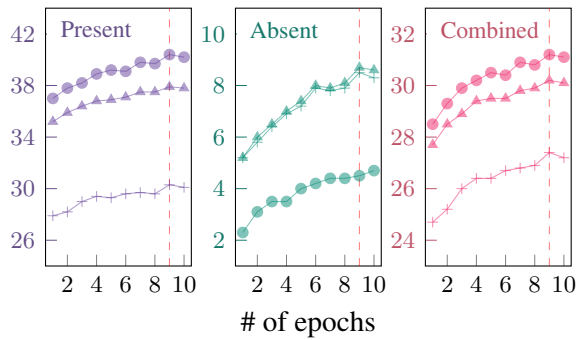
\begin{figure}[htb!]
    \centering
    \pgfplotstableread{figures/data/bart-large-dev.dat}\barttable

\begin{tikzpicture}

\begin{axis}[
    width=3.5cm,
    height=5cm,
    xmin=0.5, 
    xmax=10.5,
    ymin=24,
    ymax=44,
    xtick pos=bottom,
    ytick pos=left, 
    x tick label style={font=\small},
    y tick label style={font=\small},
    xlabel={\phantom{\# of epochs}},
    ytick={26,30,...,42},
    yticklabel style=RoyalPurple!75!black,
    ytick style=RoyalPurple!75!black,
    legend style={at={(0.01,0.99)},anchor=north west},
    legend style={font=\footnotesize, text=RoyalPurple!75!black, draw=none, text width=3em},
    ]

    \addlegendimage{empty legend}
    \addlegendentry{Present}

    \addplot [opacity=0.5, color=red, dashed] coordinates { (9,24) (9,44) };

    \addplot [opacity=0.5, mark=*, color=RoyalPurple] table [x=epoch,y=prefm] {\barttable};
    \addplot [opacity=0.5, mark=triangle*, color=RoyalPurple] table [x=epoch,y=pref5] {\barttable};
    \addplot [opacity=0.5, mark=+, color=RoyalPurple] table [x=epoch,y=pref10] {\barttable};

\end{axis}
\end{tikzpicture}
\hspace*{-1em}
\begin{tikzpicture}
\pgfplotsset{every axis/.style={ymin=0},
                 y axis style/.style={
                            yticklabel style=#1,
                            ylabel style=#1,
                            y axis line style=#1,
                            ytick style=#1}}
\begin{axis}[
    width=3.5cm,
    height=5cm,
    xmin=0.5, 
    xmax=10.5,
    ymin=0,
    ymin=1,
    ymax=11,
    xtick pos=bottom,
    ytick pos=left, 
    x tick label style={font=\small},
    y tick label style={font=\small},
    xlabel={\# of epochs},
    yticklabel style=PineGreen!75!black,
    ytick style=PineGreen!75!black,
    legend style={at={(0.01,0.99)},anchor=north west},
    legend style={font=\footnotesize, text=PineGreen!75!black, draw=none, text width=3em},
    ]

    \addlegendimage{empty legend}
    \addlegendentry{Absent}

    \addplot [opacity=0.5, color=red, dashed] coordinates { (9,1) (9,11) };

    \addplot [opacity=0.5, mark=*, color=PineGreen] table [x=epoch,y=absfm] {\barttable};
    \addplot [opacity=0.5, mark=triangle*, color=PineGreen] table [x=epoch,y=absf5] {\barttable};
    \addplot [opacity=0.5, mark=+, color=PineGreen] table [x=epoch,y=absf10] {\barttable};

\end{axis}
\end{tikzpicture}
\hspace*{-1em}
\begin{tikzpicture}
\pgfplotsset{every axis/.style={ymin=0},
                 y axis style/.style={
                            yticklabel style=#1,
                            ylabel style=#1,
                            y axis line style=#1,
                            ytick style=#1}}
\begin{axis}[
    width=3.5cm,
    height=5cm,
    xmin=0.5, 
    xmax=10.5,
    ymin=23,
    ymax=33,
    xtick pos=bottom,
    ytick pos=left, 
    x tick label style={font=\small},
    y tick label style={font=\small},
    xlabel={\phantom{\# of epochs}},
    yticklabel style=WildStrawberry!75!black,
    ytick style=WildStrawberry!75!black,
    legend style={at={(0.01,0.99)},anchor=north west},
    legend style={font=\footnotesize, text=WildStrawberry!75!black, draw=none, text width=3em},
    ]

    \addlegendimage{empty legend}
    \addlegendentry{Combined}

    \addplot [opacity=0.5, color=red, dashed] coordinates { (9,23) (9,33) };

    \addplot [opacity=0.5, mark=*, color=WildStrawberry] table [x=epoch,y=allfm] {\barttable};
    \addplot [opacity=0.5, mark=triangle*, color=WildStrawberry] table [x=epoch,y=allf5] {\barttable};
    \addplot [opacity=0.5, mark=+, color=WildStrawberry] table [x=epoch,y=allf10] {\barttable};

\end{axis}
\end{tikzpicture}
    \caption{Performance of our baseline model on the KP20k validation set across training epochs, measured by $F_1@M$ ($\circ$), $F_1@5$ ({\small$\triangle$}) and $F_1@10$ ({\small$+$}) for present, absent and combined keyphrases.}
    \label{fig:bart-large-dev}
\end{figure}

At test time, we evaluate the model using greedy decoding to generate the most probable keyphrases, or beam search (K=20) to assemble the top-$k$ keyphrases across all beams.
To select the best-performing model, we save a checkpoint at the end of each epoch and evaluate its performance on the validation set of KP20k, using $F_1@\{M, 5, 10\}$ scores against the ground truth keyphrases.
Overall, fine-tuning for 9 epochs produces the highest scores (see Figure~\ref{fig:bart-large-dev}), leading us to select the corresponding checkpoint as our baseline model.

Code for training, inference and evaluation is available on \href{https://github.com/boudinfl/kg-datasets-metrics-models}{GitHub}.
Additionally, all model weights are accessible on \href{https://huggingface.co/taln-ls2n/bart-large-kp20k}{Hugging Face}.
Implementation details are given in Appendix~\ref{subsec:implementation-details}.

\begin{table}[b!]
    \centering

\begin{tabular}{l@{\quad}l|rrrr}
\toprule
\multicolumn{2}{l}{\textbf{Metric}}  & \textbf{Ours} & \textbf{Best} & \textbf{\#~$\downarrow$} & \textbf{\#~$\uparrow$} \\
\midrule

\multirow{2}{*}{{\footnotesize $F_1@M$}} 
      & {\small Present}  & 39.9 & 45.3 & 19 & 6 \\
    ~ & {\small Absent}   & 4.5  & 11.2  & 9  & 13 \\
    
\cmidrule{1-6}

\multirow{2}{*}{{\footnotesize $F_1@5$}} 
     & {\small Present}  & 37.7 & 42.6 & 19 & 6 \\
   ~ & {\small Absent}   & 8.2  & 7.3  & 23 & 0 \\
\bottomrule
\end{tabular}




    \caption{Performance of our baseline model on the KP20k test set, compared to the best-reported scores in literature, with the number of previous models underperforming (\#~$\downarrow$) or outperforming (\#~$\uparrow$) the baseline.}
    \label{tab:baseline-performance}
\end{table}

Here, we evaluate the performance of our baseline model on KP20k test set and compare it against previously proposed models.
Table~\ref{tab:baseline-performance} summarizes the results for both present and absent keyphrase prediction.
Our model achieves strong overall performance, surpassing most prior models and achieving state-of-the-art results in absent keyphrase prediction in terms of $F_1@5$.
We believe that this level of performance establishes our baseline model as a robust point of reference for future research.





\section{Open Challenges and Discussion}
\label{sec:challenges}

We wrap up this paper by highlighting two challenges in keyphrase generation and suggesting actionable strategies to address them.
Finally, we discuss what LLMs can do for the task.


\subsection{Benchmark Datasets}

Our analysis revealed alarming levels of redundancy between the most frequently used benchmark datasets, stressing the need to deviate from the common practice of relying on the same five datasets.
Thus, the first challenge we identified is the lack of diverse, sizeable benchmark datasets for keyphrase generation.
While recent efforts have been devoted to building new datasets, they either reuse most samples from KP20k~\cite{DBLP:conf/cikm/MahataAGKPSAS22}, contain too few samples~\cite{NEURIPS2022_f8870955} or are restricted to a specific domains~\cite{houbre-etal-2022-large,boudin-aizawa-2024-unsupervised} or goals~\cite{wu-etal-2024-metakp}.

Creating a new dataset is undoubtedly difficult, as manual annotation of keyphrases is both costly and requires domain expertise.
A practical solution is to look for naturally occurring keyphrases, and scientific papers with their author-provided keywords are a well-known match.
Another common issue of existing datasets is their lack of proper document sourcing.
For instance, the documents in KP20k were collected from ``various online digital libraries'' and lack crucial metadata  such as DOIs, authorship details or licences.
Given these considerations, we suggest leveraging arXiv for creating a new dataset as it aligns with our requirements: it offers content under Creative Commons, provides a substantial volume of categorized, identified and machine-readable (\LaTeX) documents. 

\subsection{Evaluation Metrics}

The second challenge we identified, which connects to the benchmark datasets, concerns the questionable robustness of automatic evaluation.
%
There are two main issues with current evaluation methods.
First, keyphrases are task-dependent.
For example, keyphrases relevant for document indexing may differ from those relevant for reading comprehension.
This aspect is rarely addressed in previous studies, despite its significant implications, notably on the need for distinct ground truth keyphrases depending on the targeted task.
Second, commonly-used evaluation metrics rely on simple matching against a single ground truth, which is likely to be incomplete.

One potential solution to address these issues is to rely on extrinsic evaluation, that is, assessing the performance of keyphrase generation models through downstream tasks.
For instance, prior works have proposed to evaluate models through their impact on document retrieval effectiveness~\cite{boudin-etal-2020-keyphrase,boudin-gallina-2021-redefining}. 
%
%
%
Two other notable works in this direction are~\citet{https://doi.org/10.1002/asi.24749}, which evaluates keyphrases in a task-oriented setting to assist reader comprehension, and~\citet{wu-etal-2024-kpeval}, which examines the alignment between keyphrases and LLM-generated queries.
However, the additional computational costs associated with conducting such extrinsic evaluations may hinder their adaption.
Here, we suggest testing the ability of LLMs to evaluate generated keyphrases, as this approach has proven successful in several tasks~\cite{chiang-lee-2023-closer}.

\subsection{LLMs for Keyphrase Generation}

Keyphrase generation stands out as one of the few NLP tasks where LLMs have not yet replaced dedicated supervised models.
Nonetheless, initial efforts to leverage LLMs for this task, primarily using in-context learning~\cite{song2023chatgptgoodkeyphrasegenerator,martínezcruz2023chatgptvsstateoftheartmodels,10.1145/3627673.3680093}, have demonstrated promising results.
Recently, \citet{shao-etal-2024-one2set} validated the effectiveness of LLMs as a keyphrase reranking method for dedicated models.
Here, we highlight two important considerations when using LLMs for keyphrase generation.

The first is data contamination, which occurs when test data is included in the model’s training data.
Given the extensive size and diverse sources of pre-training datasets used for LLMs, it is likely that widely available documents composing the current benchmarks have been included.
Solutions to address this issue are not straightforward, but applying pre-training data detection methods~\cite{zhou-etal-2024-dpdllm,zhang-etal-2024-pretraining} to identify and mitigate data leakage is a necessary first step.

The second is the computational costs.
Generating keyphrases using LLMs across a vast collection of documents is prohibitively expensive.
While ``lightweight'' models~\cite{grattafiori2024llama3herdmodels} or fast inference strategies~\cite{liu-etal-2024-speculative-decoding} are being developed to reduce these costs, scalable solutions remain an open challenge.
Reporting the performance-inference speed trade-off of future models would help better position their practical usefulness.

\section*{Limitations}

\subsection*{Scope of the analysis}

While we are confident that the sample of papers covered in this analysis provides a comprehensive representation of the research on keyphrase generation, our selection is not exhaustive. 
Specifically, it does not account for papers published in non-ACL journals or hosted on pre-print servers, which may present additional perspectives or recent advancements in the field.
Our analysis focuses on keyphrase generation and does not cover the closely related field of keyphrase extraction, which converges on the datasets and evaluation metrics.

\subsection*{Manual extraction of best scores}

Our analysis focuses on the best scores reported for the models and could be extended to include baselines and ablation studies.
Collecting the best scores from the selected papers was not always possible due to typos or ambiguities in the tables.
Furthermore, our disambiguation strategy---selecting either the model demonstrating the best overall performance or, when unclear, the one performing best on the KP20k dataset---may result in suboptimal scores for other datasets.

\bibliography{anthology,custom}

\clearpage

\newpage

\appendix

\section{Appendix}
\label{sec:appendix}

\subsection{Related Surveys}
\label{subsec:related-surveys}

To our knowledge, this is the first attempt at compiling and analyzing the performance of keyphrase generation models.
%
In contrast, several surveys have been carried out on keyphrase extraction, starting with~\cite{hasan-ng-2014-automatic}, which focused on pre-deep-learning unsupervised methods.
%
%
Subsequent surveys, such as~\cite{10.23919/FRUCT48121.2019.8981519}, \cite{https://doi.org/10.1002/widm.1339} and \cite{Firoozeh_Nazarenko_Alizon_Daille_2020}, included additional, more recent methods and presented comparative experimental studies.
More recently, \citet{song-etal-2023-survey} carried out a comprehensive review of keyphrase extraction methods, covering PLM-based models, and \citet{XIE2023103382} performed a large-scale analysis of keyphrase prediction methods, which included results from some generative models.
Despite marked differences, notably in the model architectures and training procedures, previous research on keyphrase extraction and generation converge on the datasets and evaluation metrics, making these surveys complementary to ours.


\subsection{Statistics of the Benchmark Datasets}
\label{sec:details-datasets}

Detailed statistics of the datasets are provided in Table~\ref{tab:datasets}.

\subsection{Details of Evaluation Metrics}
\label{sec:details-eval}

For a given document $d$, the performance of a model is evaluated by comparing its predicted keyphrases $\mathcal{P}=\{p_1, p_2, \cdots, p_M\}$ with a set of gold truth keyphrases $\mathcal{Y}=\{y_1, y_2, \cdots, y_O\}$.
Keyphrases are lowercased, stemmed with the Porter Stemmer~\cite{10.5555/275537.275705}, and duplicates are removed prior to score calculation.
When only the top-$k$ predictions $\mathcal{P}_{:k}=\{p_1, \cdots, p_{\min(k, M)}\}$ are used for evaluation, the \emph{precision}, \emph{recall} and \emph{$F_1$ measure} are computed as follows:
\begin{gather*}
    P@k = \frac{|\mathcal{P}_{:k} \cap \mathcal{Y}|}{|\mathcal{P}_{:k}|} \quad
    R@k = \frac{|\mathcal{P}_{:k} \cap \mathcal{Y}|}{|\mathcal{Y}|} \\
    F_1@k = 2 \times \frac{P@k \times R@k}{P@k + R@k}
\end{gather*}
The most commonly used metrics are defined as:
\begin{itemize}
    \item $F_1@5$: $F_1@k$ when $k=5$.
    \item $F_1@10$: $F_1@k$ when $k=10$.
    \item $F_1@M$: $M$ denotes the number of predicted keyphrases. Here, all the predicted phrases are used for evaluation, i.e.~without truncation.
    \item $F_1@O$: $O$ denotes the number of gold truth keyphrases.
    \item $R@10$: $R1@k$ when $k=10$.
    \item $R@50$: $R1@k$ when $k=50$.
\end{itemize}

Noting that when using the top-$k$ predictions and the number of predicted keyphrases $M$ is lower than $k$, incorrect phrases are appended to $\mathcal{P}$ until that $M$ reaches $k$.

A keyphrase is labelled as present if it constitutes a subsequence of token of $d$ (in stemmed form), and absent otherwise.
This method is stricter than regex-based matching commonly used in previous work.
When results for present and absent are reported separately, only the present or absent keyphrases from $\mathcal{P}$ and $Y$ and used for score calculation.
Papers usually report the macro-average scores over all the data examples in a benchmark dataset.

\subsection{Example of normalized keyphrases}
\label{subsec:example-normalization}

An example of data normalization as in~\citet{meng-etal-2017-deep}\footnote{\url{https://github.com/memray/OpenNMT-kpg-release/blob/d16bf09e21521a6854ff3c7fe6eb271412914960/notebook/json_process.ipynb}} is presented in Table~\ref{tab:example-normalization}.

\begin{table}[!ht]
    \centering
    \begin{tabular}{p{.9\linewidth}}
\toprule
\textbf{Title}: Autoimmune polyendocrinopathy candidiasis ectodermal dystrophy: known and novel aspects of the syndrome \\
\cmidrule{1-1}
\textbf{Abstract}: Autoimmune polyendocrinopathy candidiasis ectodermal dystrophy (APECED) is a monogenic autosomal recessive disease caused by mutations in the autoimmune regulator (AIRE) gene and, as a syndrome, is characterized by chronic mucocutaneous candidiasis and the presentation of various autoimmune diseases. During the last decade, research on APECED and AIRE has provided immunologists with several invaluable lessons regarding tolerance and autoimmunity. This review describes the clinical and immunological features of APECED and discusses emerging alternative models to explain the pathogenesis of the disease. \\
\midrule
\textbf{Keyphrases}: apeced -- aire -- chronic mucocutaneous candidiasis -- il-17 -- il-22 \\
\textbf{Normalized}: apeced -- aire -- chronic mucocutaneous candidiasis -- il \texttt{<digit>} \\
\bottomrule
\end{tabular}

    \caption{Example of document from KP20k (S2CID: 32645143) with its associated keyphrases and their normalized forms.}
    \label{tab:example-normalization}
\end{table}

\subsection{Implementation Details}
\label{subsec:implementation-details}

We use the BART-large model weights\footnote{\url{https://huggingface.co/facebook/bart-large}} as our initial pre-trained language model and perform fine-tuning on the KP20k training set\footnote{\url{https://huggingface.co/datasets/taln-ls2n/kp20k}} for 10 epochs.
We use the AdamW optimizer with a learning rate of 1e-5 and a batch size of 4.
Fine-tuning the model using 2 Nvidia GeForce RTX 2080 took 400 hours.

\begin{table*}[ht]
    \centering
    \begin{tabular}{lr@{~/~}r@{~/~}rrrrr}
\toprule
    \textbf{Dataset} &
    \textbf{train} & \textbf{dev} & \textbf{test} & 
    \textbf{\#kp} &  \textbf{|kp|}  & 
    \textbf{\%abs}
    \\
\midrule
   
    KP20k~\small{\cite{meng-etal-2017-deep}} & 
    514k & 20k & 20k &
    5.3 & 2.1 &
    36.7 \\
    
    SemEval-2010~\small{\cite{kim-etal-2010-semeval}} &
    144 & -- & 100 &
    15.7 & 2.1  &
    55.5 \\
    
    Inspec~\small{\cite{hulth-2003-improved}} &
    1k & 500 & 500 &
    9.6 &  2.3 &
    21.5 \\

    Krapivin~\small{\cite{krapivin2009large}} & 
    1844 & - & 460 &
    5.2 & 2.2 &
    43.8 \\
    
    NUS~\small{\cite{10.1007/978-3-540-77094-7_41}} &
    -- & -- & 211 &
    11.5 & 2.2 & 
    48.7 \\

    DUC2001~\small{\cite{10.5555/1620163.1620205}} &
    -- & -- & 308 &
    8.1 & 2.1 & 
    2.7 \\
    
    KPTimes~\small{\cite{gallina-etal-2019-kptimes}} &
    260k & 10k & 20k &
    5.0  & 1.5  & 54.4 \\

    StackEx~\small{\cite{yuan-etal-2020-one}} &
    298k & 16k & 16k &
    2.7 & -- &
    42.5 \\

    Weibo~\small{\cite{wang-etal-2019-topic-aware}} &
    37k & 4.6k & 4.6k & 
    1.1 & 2.6 &
    75.8 \\
    
    StackEx~\small{\cite{wang-etal-2019-topic-aware}} &
    39.6k & 4.9k & 4.9k &
    2.4 & 1.4 &
    54.3 \\

\bottomrule
\end{tabular}









    \caption{Statistics of the benchmark datasets taken from~\cite{10.5555/1620163.1620205,gallina-etal-2019-kptimes,wang-etal-2019-topic-aware,yuan-etal-2020-one,do-etal-2023-unsupervised}}
    \label{tab:datasets}
\end{table*}







\end{document}